\documentstyle[epsfig]{aipproc}
\begin{document}
\title{Origin of Cosmic Radiation}
\author{Thomas K. Gaisser\thanks{Research supported in part by
the U.S. Department of Energy under Grant DE-FG02-91ER40626}
}
\address{Bartol Research Institute, University of Delaware\\
Newark, DE 19716}

%%%%%%%%%%%%%%%%%%%%%%%%%%%%%%%%%%%%%%%%%%%%%%%%%%%%%%%%%%%%%%%%%%%%%%%%

\maketitle
\begin{abstract}
I give a brief overview of cosmic ray physics,
highlighting some key questions and how they will be
addressed by new experiments.
\end{abstract}

%%%%%%%%%%%%%%%%%%%%%%%%%%%%%%%%%%%%%%%%%%%%%%%%%%%%%%%%%%%%%%%%%%%%%%%%
\section{Introduction}
There are several new experiments in cosmic-ray physics
and related fields running now or planned for the near
future.  These include measurements of antiprotons with
balloons and from space, new
gamma-ray detectors to cover the full energy range from
$<100$~MeV to $>10$~TeV, studies of the knee region
with a variety of experiments, giant air
shower detectors for the highest energy particles, and
huge, deep neutrino detectors to find astrophysical
sources of high energy neutrinos.  In this talk, I will
review the key questions driving these experiments and,
where possible, put them in a larger context and relate 
them to each other.

After some brief remarks about the sun as a cosmic-ray
source, I review cosmic rays of galactic origin and then
extragalactic cosmic rays.  Energy content of the
various components of the cosmic radiation and the
corresponding power to maintain the observed
fluxes are used as a guide to possible sources.

\section{Solar cosmic rays, a paradigm}

One class of solar cosmic rays consists of solar
energetic particles accelerated during solar
flares~\cite{Rufolo}.  The high energy accelerated particles 
can be detected with neutron monitors~\cite{JRyan}.
Gamma-radiation produced by interactions
of the accelerated particles in the solar atmosphere
are also seen, including the contribution from
$\pi^0\rightarrow\gamma\gamma$~\cite{Ranketal}.
Therefore we can infer
that secondary neutrinos must also be present.
In this sense the sun is an ideal cosmic ray source,
in that both the accelerated particles and the secondaries from
their interactions at the source are directly identified
and associated with the particular accelerator.

The large coronal mass ejections associated with some
solar flares drive shocks into the interplanetary
medium at which particles are energized by diffusive
shock acceleration. This is but one example of
acceleration by shocks in the heliosphere. Energetic
particles are typically observed in spatial association
with shocks as they pass by a spacecraft such as 
ACE~\cite{ACE}, concentrated particularly in the 
downstream region.

The anomalous cosmic rays are a distinct subset of
particles accelerated in the heliosphere, which are
now understood to be particles of interstellar origin
accelerated at the termination shock in the solar 
wind~\cite{anomalous}.  This phenomenon reminds
us that cosmic-ray acceleration typically occurs
in regions with structure created by strong stellar
winds and past supernovae.  An important analog for galactic cosmic rays
is the wind-blown bubble surrounding a massive star.
The {\it astropause}, by analogy with the heliopause, is
the contact discontinuity between the fast
stellar wind and the surrounding medium.  There is both
a bow shock outside the astropause and a termination
shock inside~\cite{Lozinskaya}.  
An interesting possibility is to consider
the signatures for acceleration by a
supernova blast wave propagating through this
kind of environment~\cite{Berezhko}.  In addition,
acceleration at multiple shock structures may have 
significant effects, both as a way of achieving
higher energy~\cite{Axford,Klepach}
and for modifying the low energy spectral shape~\cite{Klepach}
for particles injected after acceleration by individual shocks.
 
\begin{figure}[htb]
\flushleft{\epsfig{figure=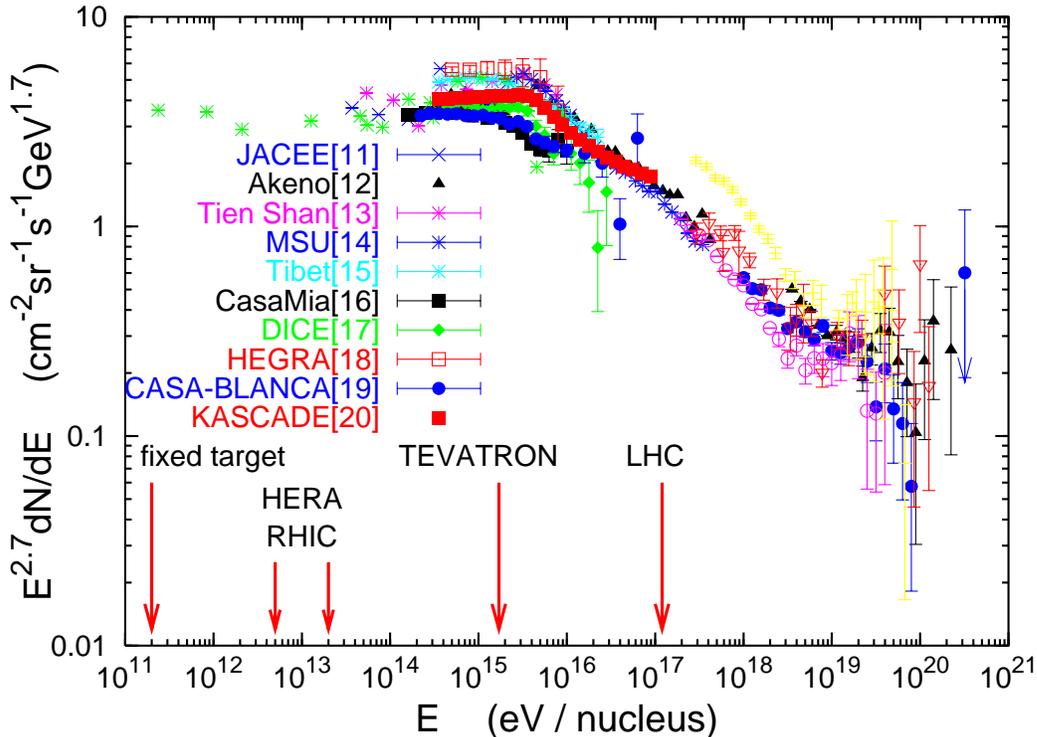,width=14cm}}
\caption{Summary of measurements of the high energy cosmic-ray spectrum.
Data below 100 TeV are from a satellite detector~\protect\cite{Grigorov}.
Data in the knee region are from
Refs.~\protect\cite{JACEE,Akeno,Dani,Fomin,Tibet,Glas,DICE,HEGRA,CasaBlanca,Kascade}.
References for 
data above \protect$10^{17}$~eV are given in Fig.~\protect\ref{Fig4}.}
\label{Fig1}
\end{figure}

\section{Galactic cosmic rays}

While solar cosmic rays are identified by their temporal
association with solar flares or spatial association with
interplanetary shocks, or by the compositional
and spectral signatures of anomalous cosmic rays,
the steady, nearly  isotropic flux of high energy
particles comes from sources far outside the heliosphere.
These sources still lack definitive identification
nearly a century after their discovery.  The fundamental
difficulty is that diffusive propagation in the turbulent
interstellar medium smooths out spatial and temporal variations
that may characterize the sources.

Fig.~\ref{Fig1} shows
a global view of the cosmic-ray spectrum.  This is
the differential ``all-particle'' spectrum (flux of
all particles summed over all species as a function
of total energy  per nucleus) multiplied by $E^{2.7}$.
Because of the low intensity at high energy, information
from direct measurements made with balloons and spacecraft
above the atmosphere is sparse for $E_0>100$~TeV.  This
is therefore
the realm of indirect extensive air shower experiments.
The equivalent laboratory energies of several major
terrestrial accelerators are noted as a reminder that
interpretation of measurements of air showers depends
on models of hadronic interactions to
extrapolate beyond the reach of laboratory measurements.
0 of high energy cascades in the atmosphere
is governed by properties of interactions of pions
and nucleons with nuclei in the atmosphere, primarily
in the forward fragmentation region (i.e. the region
of phase space in which the secondaries carry a 
non-vanishing fraction of the incident energy  as 
$E_0$ increases). In contrast, the highest energy
collider experiments study primarily the central
region of phase space for proton-proton (antiproton)
collisions.
The two features in the spectrum as shown in Fig.~\ref{Fig1}
are the {\it knee} between $10^{15}$ and $10^{16}$~eV
and the {\it ankle} around $10^{19}$~eV, which will
be discussed below.

\subsection{Low energy galactic cosmic rays}

\begin{figure}[htb]
\flushleft{\epsfig{figure=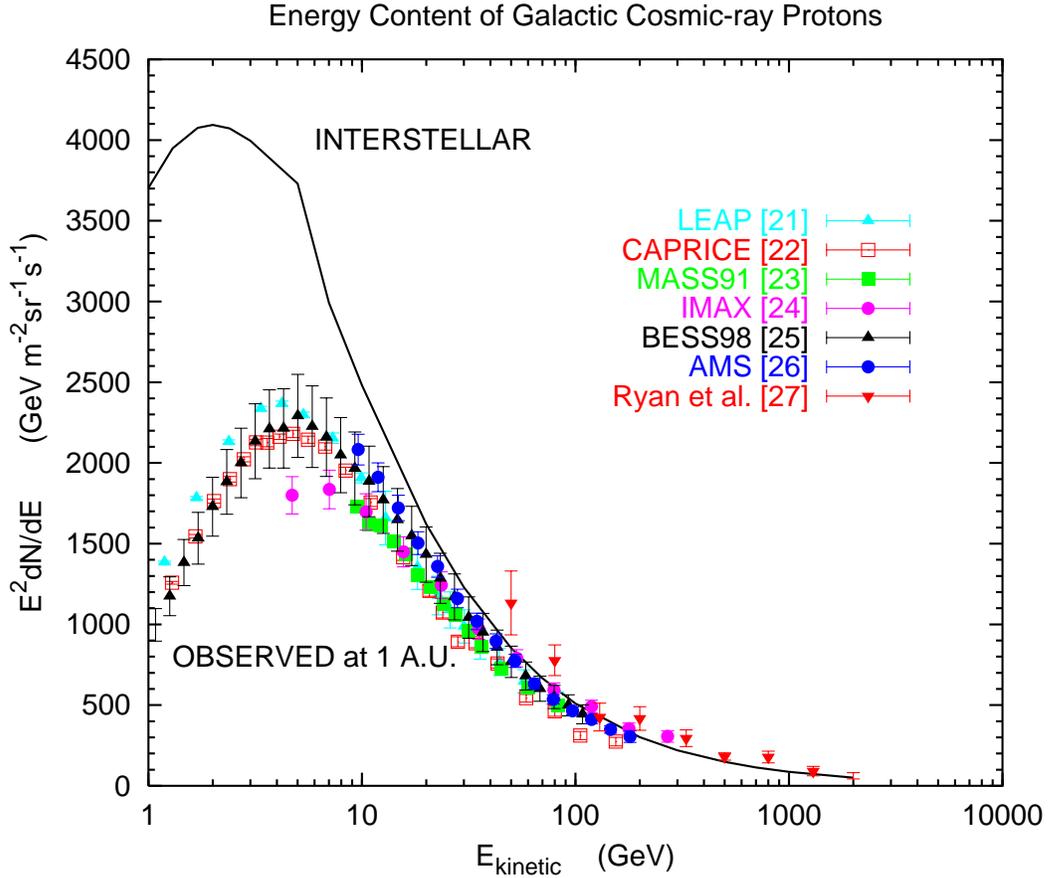,width=14cm}}
\caption{Summary of direct measurements proton spectra
made with detectors on balloons and in space.}
\label{Fig2}
\end{figure}

There are several recent, independent measurements
of the primary proton and helium spectra 
for $E<100$~GeV made with
balloon-borne 
spectrometers~\cite{LEAP,CAPRICE,MASS91,IMAX,BESS98}.
Fig.~\ref{Fig2} is a compilation
of proton spectra from these measurements as well as
the newly published data from the test flight of
the AMS detector on the Space Shuttle~\cite{AMSprotons}.
At higher energy into the TeV region the principal
data is still that of the calorimeter measurement
of Ryan, Ormes and Balasubrahmanyan~\cite{Ryan}.
The differential spectra are multiplied by $E^2$
so that equal areas on the semilogarithmic plot correspond
to equal contributions to the energy content of 
the cosmic radiation.  

Also shown in Fig.~\ref{Fig2} (by the line) is an estimate of
the interstellar spectrum after accounting for the
effects of solar modulation, which
suppresses the intensity of low-energy
particles in the inner heliosphere~\cite{modulation}.
The median energy in the local interstellar medium (ISM) 
is about 6 GeV, and more than
90\% of the energy is carried by particles with
$E<50$~GeV.  The energy density in cosmic rays in
the local ISM is
\begin{equation}
P_{CR}\;=\;\rho_{E,CR}\;=\;{4\pi\over c}\int\,E\,{dN\over dE}\,dE\;\sim\;
10^{-12}\;\;{\rm erg/cm}^3.
\label{Edensity}
\end{equation} 
Assuming this is typical of the entire galactic disk,
the power required to maintain this cosmic-ray energy
density in steady state can be estimated as
\begin{equation}
{\rho_{E,CR}\,V_{disk}\over \tau_{esc}}\;\sim\;
{10^{-12}\times 10^{67}\over 10^{14}}\;\sim\;
10^{41}\,{{\rm erg}\over{\rm s}},
\label{CRpower}
\end{equation}
where $\tau_{esc}$ is the average time spent
by cosmic rays in the disk of the Galaxy before they
are lost into intergalactic space.
The similarity of this power requirement to
the power available in kinetic energy of supernova
ejecta ($10^{51}$~erg/30~yrs~$\sim10^{42}$~erg/s)
is one of the principal arguments for supernova
explosions as the source of cosmic rays~\cite{Ginzburg}.

The characteristic time in Eq.~\ref{CRpower} is an
effective parameter that emerges from
a range of more realistic models for diffusion
and propagation of cosmic rays in the disk and halo of
the Galaxy~\cite{Ptuskin}.  In general $\tau_{esc}$
depends on energy.  The relation between the source
spectrum, $Q(E)$ and the observed spectrum is
\begin{equation}
\tau_{esc}(E)\,Q_{source}(E)\;\propto\;
{dN\over dE}_{observed}\;\propto\;E^{-\alpha}.
\label{source-spectrum}
\end{equation}
The observed differential spectral index is
$\alpha\,\approx\,2.7$ for protons and primary nuclei
such as carbon, oxygen, etc., with small but interesting
differences among the various nuclei~\cite{Swordy}.

The energy dependence of $\tau_{esc}$ can be measured by
comparing the spectrum of a secondary nucleus to that of
a parent primary nucleus (e.g. B/C 
or sub-Fe/Fe).  Primary nuclei are those
accelerated in the sources.  Secondary nuclei are
those essentially absent in the source but which
can be produced by spallation of primary nuclei of
larger mass number.  As a consequence, their spectral
index at production is known.  To the extent that
spallation cross sections
are approximately constant in energy, the spectral
index at production for a secondary nucleus is the 
{\em observed} spectral index of its parent.  
Thus the analog of Eq.~\ref{source-spectrum} for a
secondary nucleus is
\begin{equation}
\tau_{esc}(E)\times {dN\over dE}_{observed}\;\propto\;{dN\over dE}_{secondary}.
\end{equation}
The data on the secondary to primary ratio
up to 20--50~GeV may be fit as a power law with the result,
\begin{equation}
\tau_{esc}(E)\;\propto\;E^{-\delta},
\end{equation}
with $\delta\sim 0.6$~\cite{Garcia-Munoz}.

The energy-dependence of $\tau_{esc}$ must be
taken into account in estimating the power requirement
for cosmic rays.  Since $\tau_{esc}(E)\,\propto\,E^{-\delta}$, it follows
from Eq.~\ref{source-spectrum} that $Q_{source}(E)\,\propto\,E^{-\gamma}$
with $\gamma\,=\,\alpha-\delta\,\approx\,2.1$.
Thus the source spectrum is harder than the
observed spectrum, and a more correct version of Eq.~\ref{CRpower} is
\begin{equation}
P_{CR}\;=\;{4\pi\over c}\,\int\,E\,Q_{source}(E)dE\;=\;
{4\pi\over c}\,\int\,E\,
{dN\over dE}_{observed}\times{1\over\tau(E)}\,dE.
\label{source-power}
\end{equation}
This relation between observed flux and source power
is true in general for other components of the cosmic
radiation and will be used below in other contexts.

The basic picture described above
has been known and understood for a long time.  
A recent result relevant to cosmic-ray propagation
in the Galaxy
is the measurement with high statistics of the spectrum of
antiprotons from $\sim 300$~MeV to $\sim 3$~GeV with the
BESS detector~\cite{BESS}.  It is an interesting test
of the model because most of the antiprotons are
produced by cosmic-ray protons, which could in principle
have a completely different distribution of sources
from the heavier nuclei like carbon, oxygen and iron
on which the model is based.  Instead, the measurements
are quite consistent with the propagation parameters
determined from the ratios of secondary to primary
nuclei~\cite{Bieberetal}.

The canonical tracer of
cosmic-ray propagation is the production of gamma-radiation
by interactions of cosmic rays in the ISM~\cite{Hunter},
as reviewed by Hunter at this conference.  
The EGRET data~\cite{Hunter} fits the general picture 
well with, however, the exception 
that the spectral index above a GeV 
is harder than expected.  The GeV range for secondary
photons is sufficiently high
that the relevant production cross sections approximately scale with
energy.  As a consequence, at such energies the
production spectrum of the photons from decay of $\pi^0$
produced by cosmic-ray interactions in the ISM should have
the same spectral index as the parent nucleons 
(mostly protons).  Since the photons propagate directly,
their observed spectral index should also be 
$\alpha\,\approx\, 2.7$.  Instead it is more
like $\alpha_\gamma\,\approx\,2.4$.
An important possibility to consider is that
there may be some contribution of interactions
at the source with a characteristically harder
spectrum.\cite{Cowsik,Berezhko2}

\subsection{Supernova model of cosmic-ray origin}

If supernovae are the sources of galactic cosmic rays
capable of accelerating particles to $\sim10^{15}$~eV
or higher, then {\em at some level} they should also
be point sources of $\gamma$-rays produced by 
interactions of the accelerated particles in or near
the source~\cite{Drury}.
The intensity depends, however, on the
degree of mixing between the high energy particles
and the gas nearby.  Certain supernova remnants (SNR)
may be sources of GeV photons~\cite{Esposito},
but they have not been detected by the TeV air Cherenkov
telescopes at the levels that would be expected
if the parent proton spectrum extended to $\sim100$~TeV
and the source spectrum were hard ($\alpha\approx2.1$)~\cite{Buckley}.

Theoretical calculations of first order diffusive
acceleration at shocks driven by supernova blast waves
show that the spectral index should be approximately
$\alpha\approx2.1$, not far from the ideal value
of $2.0$ for test particle acceleration at strong 
shocks~\cite{Berezhko3}.  The calculations include the
non-linear effect of back-reaction of the cosmic-rays
on the shock. The conclusion is that a large fraction
($>10$\%) of the kinetic energy of the supernova
remnant is deposited in accelerated particles
and that the spectrum extends to $E_{max}$ approaching $10^{15}$~eV.

There are several possibilities for explaining the absence of
TeV $\gamma$-rays at the level expected by using a hard spectrum
to extrapolate the GeV observations of Ref.~\cite{Esposito}.
It is possible~\cite{Gaisseretal} that the spectrum of accelerated
particles in the source is significantly steeper than
($\alpha\sim 2.4$), in which case the extrapolation to 
$E_\gamma\,>$\,TeV of fits to $\sim$GeV
data can be consistent with the
upper bounds from air Cherenkov telescopes.
Alternatively, the observed photons in these sources may be
not be associated with acceleration of cosmic rays or the maximum
cosmic-ray energy may be low in these particular remnants.

TeV gamma-rays have now been detected from the supernova remnants
SN1006~\cite{SN1006} and CasA~\cite{CasA}, and the potential importance
of inverse Compton scattering by high energy electrons
is emphasized~\cite{Pohl,Mast}.  The contribution of
$\pi^0\rightarrow\gamma\gamma$ associated with cosmic ray acceleration
is still uncertain.  (For a review see Ref.~\cite{Baring}.)

\subsection{The {\em knee} of the spectrum}

One interpretation of the knee of the spectrum
is that it reflects a change in propagation of
galactic cosmic rays, perhaps corresponding to more rapid
escape from the galaxy (effectively an increase in
$\delta$)~\cite{Pt1}.  A problem for this interpretation
is that the spectrum in the knee region may have more structure
than would be the case for a steepening of the rigidity
spectrum of each elemental component of the cosmic radiation.
This can be seen best by following one of the individual 
data sets through the bend above $10^{15}$~eV in Fig.~\ref{Fig1},
but there are significant differences among the different measurements.
An alternate interpretation is that
this part of the spectrum may be produced by
only one or a few sources~\cite{Erlykin}.

Another problem is that simple extrapolation of
$\tau_{esc}(E)\propto E^{-\delta}$ with $\delta\approx0.6$
breaks down around $3\times10^{15}$eV since
the effective escape length $c\tau_{esc}\sim 300$~pc
at that energy.  This is corresponds to just one
crossing of the disk and would tend to produce a
large and increasing
anisotropy approaching this energy.  The measured anisotropy
does increase around $10^{15}$~eV~\cite{Erlykin,Clay},
but even at high energy the amplitude of the first
harmonic is only a few percent.  For $E<10^{15}$~eV it is
$\sim10^{-3}$ or somewhat less~\cite{Munakata}.
It has been noted~\cite{SeoPtuskin,Simon} that a Kolmogorov spectrum
of turbulence in the interstellar medium indeed
predicts that the diffusion coefficient should
increase like $E^{1\over 3}$, which corresponds
to $\delta={1\over 3}$ rather than $0.6$.  The compilation of
low energy measurements of the ratios of 
secondary to primary nuclei~\cite{Garcia-Munoz}
can be made consistent
with this by invoking reacceleration of the
injected spectrum in the ISM~\cite{SeoPtuskin,Simon}.
Note that  the source spectral index in this scenario
would need to be larger ($\gamma\approx 2.4$)
to maintain $\alpha\,=\,\gamma\,+\,\delta\,\approx 2.7$
at high energy, in contradiction to the calculations
of diffusive shock acceleration in shock waves
from SNR mentioned above~\cite{Berezhko3}.  
Measurement of the ratio of boron to carbon can be
fit well with a hard source spectrum and $\delta\approx0.6$~\cite{Swordy2}.
The highest energy point of Ref.~\cite{Swordy2} is about two $\sigma$
below the fit~\cite{SeoPtuskin} that uses reacceleration and $\delta=0.33$.
Measurements of the secondary/primary ratios at higher energy are needed.

\begin{figure}[htb]
\flushleft{\epsfig{figure=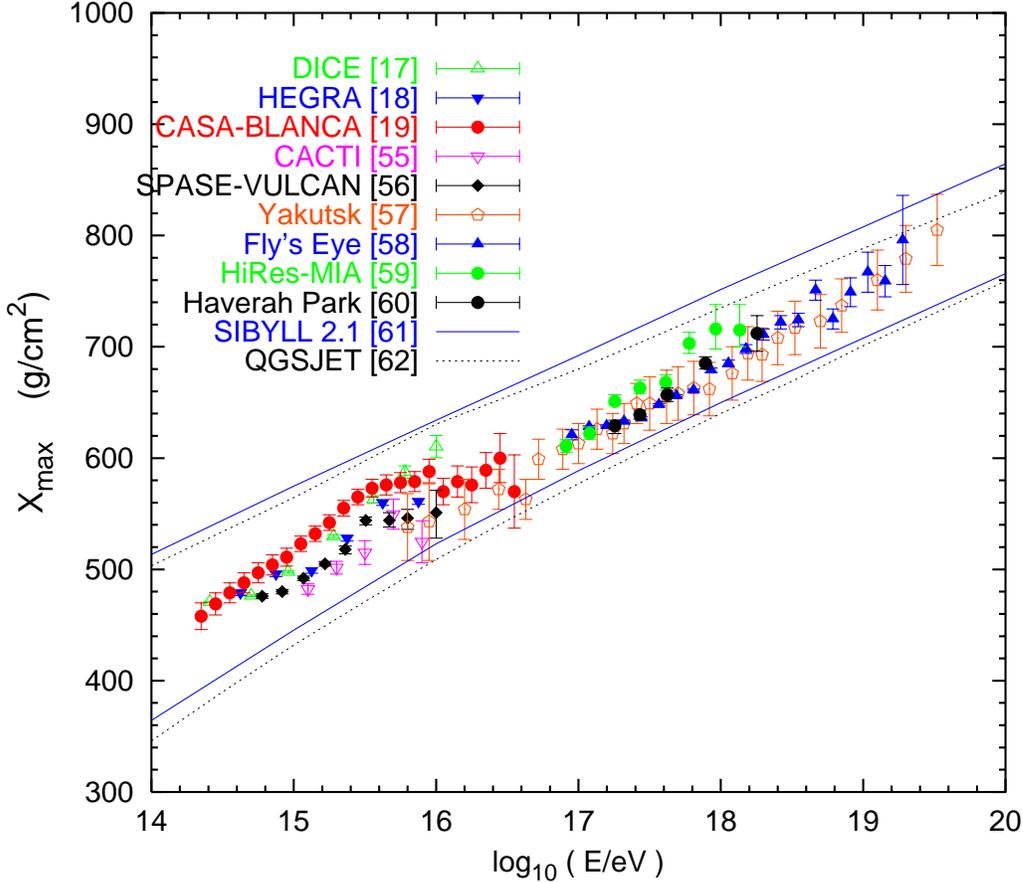,width=14cm}}
\caption{Summary of measurements of average depth of shower maximum vs. 
primary energy.  Plots of model curves over the full range of
energy are given in Ref.~\protect\cite{Pryke}.}
\label{Fig3}
\end{figure}

An important diagnostic of the knee region is the way in which
the average depth of shower maximum ($X_{max}$) depends on energy.
($X_{max}$ is defined as the depth along the shower axis at which
the shower has maximum size or number of charged particles.)
The most direct (mass independent)
measurements of $X_{max}$ are made with 
optical detectors, which collect light over a significant
portion of the shower, including the maximum.
One approach is to use the Cherenkov light projected
around the shower axis as it develops through the
atmosphere~\cite{DICE,HEGRA,CasaBlanca,Cacti,Vulcan,Yakutsk}.
At sufficiently high energy, fluorescence light can be used
to map out the shower profile and hence determine depth of shower 
maximum~\cite{Birdetal}.  The intermediate energy range is probed
by the hybrid HiRes/MIA coincidence experiment~\cite{MIAHiRes} and by
indirect methods.~\cite{Hinton}.  There is currently much
activity in this field, as summarized in Fig.~\ref{Fig3}.

Calculations are needed to interpret the data. 
Results of two calculations~\cite{SIBYLL21,QGSjet}
are shown in Fig.~\ref{Fig3}.  The upper pair of lines is for showers
initiated by protons and the lower pair for iron-initiated showers.
Note the tendency, most clearly visible 
in Ref.~\cite{CasaBlanca}, for a transition from heavies
to a larger fraction of protons above $10^{15}$~eV,
followed by a transition  back toward heavies above
the knee, around $10^{16}$~eV.  There are many systematic
uncertainties, so this can only be considered suggestive, 
but it would fit in with the suggestion~\cite{Erlykin}
of one or two specific sources
contributing in the knee region.  The increase of protons
just above $10^{15}$~eV would be from the high energy source
with the heavy nuclei of the same rigidity showing up in the
all-particle spectrum at higher energy.

\section{Extragalactic cosmic rays}

If the cosmic rays in the knee region and above were
all of extragalactic origin, as has been suggested
from time to time (e.g.~\cite{RJP}) then
the isotropy in the knee region would no longer
be a problem, assuming particles of this energy
could diffuse into the galaxy sufficiently easily.
There is at least one piece of evidence, however,
that the cosmic radiation
below $3\times10^{18}$~eV is indeed of galactic origin.
This is the anisotropy reported by AGASA~\cite{Teshima}.
There is an enhancement of particles in the energy
bin around $10^{18}$~eV from near the galactic center.
This anisotropy disappears rather quickly at higher
energy.  Data from Fly's Eye~\cite{FEanisotropy} and
the SUGAR array~\cite{Clay2} show a similar behavior.

\subsection{Transition from Galactic to extragalactic}
Somewhere around $10^{18}$~eV or a bit higher, one indeed
expects a transition to cosmic rays of extragalactic
origin because the gyroradius in a $3\,\mu$Gauss galactic
field is comparable to kpc galactic scales.  
The Fly's Eye presented circumstantial
evidence for such a transition
from their stereo data~\cite{Birdetal}.
It consists of an apparent transition from 
heavy to light primary nuclei coupled with a hardening of the spectrum,
both around $3\times 10^{18}$~eV.  The transition from
heavy to light with increasing energy is also
present in the recent coincidence data from the HiRes
prototype and the CASA-MIA detectors~\cite{MIAHiRes},
although it seems to appear at lower energy than in
the original Fly's Eye stereo data.

The stereo Fly's Eye result
was challenged by Akeno~\cite{AkenoComp},
who reported no evidence for a transition
based on their measurements of the ratio of low
energy muons to electrons in the showers.
The analyses of the two groups are based on different
measurements and different calculations.  
Dawson {\it et al.}~\cite{Dawson} compare the two analyses
in a common framework and find some weaker but
consistent evidence for a transition.
The transition from heavier
toward lighter composition is also model-dependent,
as can be seen from Fig.~\ref{Fig3} above $10^{18}$~eV.  

\begin{figure}[htb]
\flushleft{\epsfig{figure=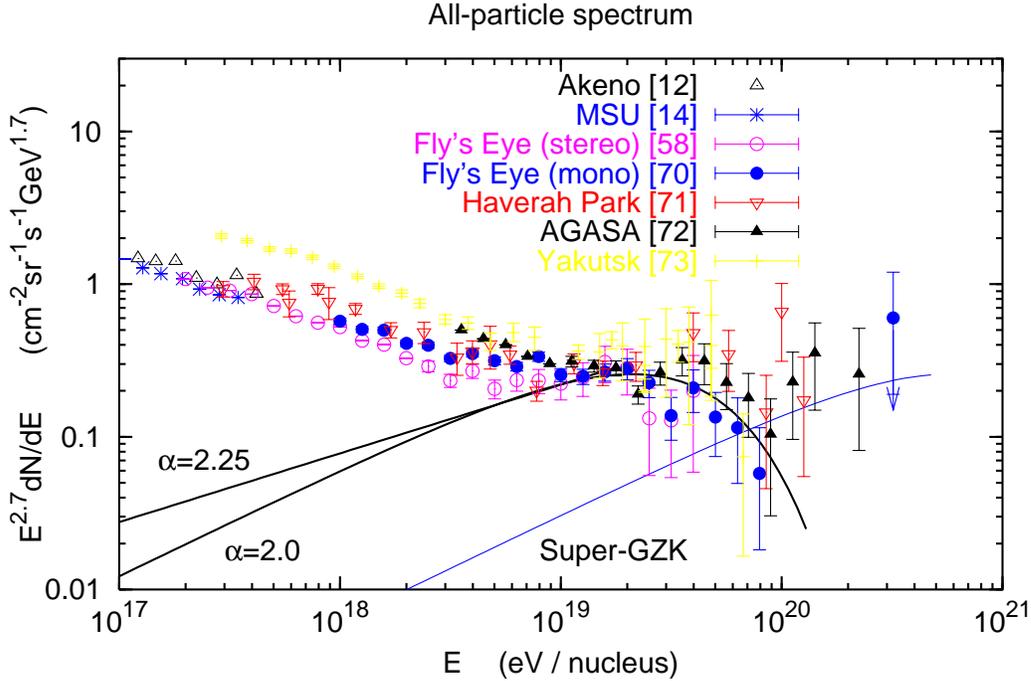,width=14cm}}
\caption{Summary of measurements of the cosmic-ray spectrum
above \protect$10^{17}$~eV.}
\label{Fig4}
\end{figure}

An expanded view of the highest energy part of the
spectrum is shown in Fig.~\ref{Fig4}.  The Fly's Eye stereo
spectrum~\cite{Birdetal,FE} most clearly shows a hardening
around $3\times10^{18}$~eV.  It could be that this is
a consequence of a difference in energy resolution
among the different experiments.  Energy resolution
should in principle be best for stereo Fly's Eye,
as compared to monocular Fly's Eye~\cite{FE}
or ground based experiments~\cite{HP,AGASA,Yakutsk2}.

\subsection{Models of extragalactic cosmic rays}
One approach to sources of the highest energy cosmic rays
is to examine the energy content of the observed cosmic rays
and  estimate the power needed
to supply them.   The idea is to proceed by analogy with the analysis for
galactic cosmic rays, as in Eq.~\ref{source-power}.  I have discussed
this in detail elsewhere~\cite{Puebla}.

Briefly, the curves labelled $\alpha = 2.0$ and $\alpha = 2.25$ in
Fig.~\ref{Fig4} are estimates of the extragalactic component of
the observed cosmic radiation.  The sources are assumed to be
distributed over cosmological distances, which gives rise to
a cutoff caused by energy loss due to photo-pion production.
The energy content estimated for this diffuse component depends on
the assumed spectral index.
If $\alpha = 2.0$ is assumed for the 0 to low energy, then 
the local energy density in this diffuse, extragalactic component is
$\rho_{CR}\sim 2\times 10^{-19}$~erg/cm$^3$.
This estimate assumes that the sources
accelerate a spectrum of particles from $E_{min}~\sim1$~GeV 
to $E_{max}\sim10^{20}$~eV.
An estimate of the
source power needed to produce this is then obtained, dividing
by the Hubble time, as $\sim 10^{37}$~erg/s/Mpc$^3$.  
If the spectrum is steeper (as for relativistic shocks with 
$\alpha\sim2.25$ ~\cite{Achterberg,Ostrowski}), then the power required 
is correspondingly higher (although in the case of acceleration
at relativistic shocks $E_{min}$ may be large~\cite{Vietri}).

Comparing the power estimate of $\sim 10^{37}$~erg/s/Mpc$^3$
to observed densities of
galaxies, clusters of galaxies and active galactic nuclei (AGN), 
and to rates of gamma-ray bursts (GRB), one finds in each case
that the power required per source is comparable to the power
observed from that class of object in electromagnetic radiation.
For example, a density of $10^{-7}$~AGN/Mpc$^3$ corresponds to
a power requirement of $10^{44}$ erg/s/AGN in high energy cosmic rays.
A rate of 300 GRB/yr would require $10^{53}$~erg/GRB.

\subsection{Highest energy cosmic rays}

Particles with energies of $10^{20}$~eV and above must be
from relatively nearby.  The attenuation length due to interaction
with the microwave background ($\lambda_{2.7^\circ}$) decreases with energy.
For example, $R_{50\%}$~\cite{TSS} is $100$~Mpc for a proton injected
at $10^{20}$~eV but only $20$~Mpc at $2\times 10^{20}$~eV. 
One possibility is a cosmologically local overdensity of
the same sources that produce all extragalactic cosmic 
rays~\cite{Grigorieva,BW}.  The line labelled ``Super-GZK'' in
Fig.~\ref{Fig4} can be used to estimate the energy density in
the nearby component as $4\times10^{-20}$~erg/cm$^3$, but the relevant
characteristic time $\tau(E)\sim \lambda_{2.7^\circ}/c$ decreases
with energy.  Thus the power requirement from Eq.~\ref{source-power} 
is significantly higher than an estimate based on the Hubble time.
Such a large local overdensity of sources is difficult to reconcile
with the observed distribution of galaxies~\cite{Blasi}.

There are at least two alternative possibilities.
One is that the highest energy cosmic rays are from nearby objects,
for example from young pulsars in the our galaxy, including
the halo~\cite{Olinto}
or from rapidly spinning magnetars in galaxies within the GZK
radius~\cite{dalpino}.  Another
is exotic sources, such as topological defects~\cite{TD,Bhat}
or decaying massive relics~\cite{Berezinsky}.  Exotic sources
produce parton cascades starting at some mass scale much higher
than the ultra-high energy cosmic rays.  The ratio of photons
and neutrinos to protons at such sources would be large because
most of the hadronization is to pions.  For distant topological
defects, photons would cascade down to lower energy (which also
puts constraints on the models~\cite{ProtherStanev,Lee}),
but for the decaying massive relics in the galactic halo, 
the observed Super-GZK
events themselves would have to be generated largely by
photons.  Indications are that this is not the case~\cite{HSV,Ave}.

\section{Outlook: new experiments}

There is currently a great deal of experimental activity underway
or planned that is motivated by questions raised by recent
results.  By way of summary, I briefly list some of
the new experiments and the questions
they will address. 

\subsection{Galactic cosmic rays}

One focus of the new gamma-ray detectors~\cite{Ong,Hoffman}
is to close the gap that now exists around 100 GeV between space
telescopes and ground-based detectors.  Another is to accumulate data
and find new sources of high-energy transients.  Both of
these quests are of great importance for identifying potential
sources of extragalactic cosmic rays.  In the process, GLAST
should extend the measurement of the diffuse flux from the Galaxy
and help clarify the reason for the hardness of the diffuse spectrum
observed with EGRET~\cite{Hunter}.
Another challenge for gamma-ray telescopes is to increase the
sensitivity and resolution in the study of supernova remnants
to resolve the question of whether there is production of secondary
pions associated with acceleration of cosmic-ray protons at
appropriate levels for the source.

Experiments on balloons and in space (e.g. AMS~\cite{AMSproposal} on Space
Station and
PAMELA~\cite{PAMELA} as a free-flyer) will
extend direct measurements of the primary spectra of hydrogen,
helium and other nuclei beyond 100 GeV.  Precise measurement
of these primary spectra into the multi-TeV region is needed for
better understanding of the expectations for atmospheric neutrinos,
especially for neutrino-induced, upward muons.  Another important goal
of these detectors is to extend the measurement of antiprotons
to higher energy, both as a probe of cosmic-ray propagation and
to search for an exotic contribution to the antiproton
flux~\cite{Bergstrom}.   They will also provide sensitive
searches for anti-nuclei.

At higher energy, the ACCESS~\cite{ACCESS} experiment under
development for deployment on Space Station will extend
composition measurements 
to approaching the knee of the spectrum.  An interesting
question is whether changes in composition characteristic of contributions
from individual sources will appear.  This detector should also have
the capability to extend the measurement of secondary/primary
nuclei to significantly higher energy than at present.  This will
address the problem of cosmic-ray propagation in the galaxy and
how the source spectrum is related to the observed spectrum.  Is
the source spectrum really a power law with differential spectral
index $\alpha\approx 2.1$, as expected from theory of diffusive
shock acceleration?

\subsection{Extragalactic cosmic rays}

The principal experimental goal is to increase the aperture for
large events in order to accumulate sufficient statistics to clarify
the origin of the highest energy cosmic rays.  The AGASA experiment
accumulates events with $E>10^{20}$~eV nominal energy at a rate
of one to two per year.  Based on the spectrum measured by
AGASA~\cite{AGASA} the corresponding rate for the Hi-Res Fly's
Eye~\cite{HiRes},
which is now running, should be $\approx 8$~events/yr, after
taking account of the duty factor for the fluorescence technique,
which is of order 10\%.  The Auger detector~\cite{Auger},
now under construction in
Argentina with an aperture of $\approx7000$~km$^2$sr,
would increase this to approximately 60/yr.  The proposed
Telescope Array of fluorescence telescopes~\cite{TA} would have
a similar effective aperture and rate, as would the proposed
Northern Auger detector.

Another objective for the stereo Hi-Res Fly's Eye experiment
will be to study the spectrum over a long range of energy
(from $<10^{18}$ to $>10^{20}$~eV) to clarify the evidence
for a transition from a galactic to an extra-galactic population
of cosmic rays.

The next big step in increased aperture for detection of giant air
showers would come from the proposed fluorescence detectors in space.
The Extreme Universe Space Observatory~\cite{EUSO}
is under study for deployment on the Space Station.  It would
aim for event rates an order of magnitude higher
than the largest ground-based detectors above $10^{20}$~eV.
A free-flying OWL
detector~\cite{OWL} with an optimized orbit could expect to
accumulate events at an even larger rate.

High-energy gamma-ray emitters such as AGN and GRB may also be
sources of ultra-high energy cosmic rays.  If so, then they might
be sources of high energy neutrinos as well.  In the simplest case
the relation would be through production of pions by hadronic interactions
in or near the source.
A discussion of the possible relationship between extragalactic
cosmic rays and $\stackrel{>}{\sim}$~TeV 
neutrinos is presented elsewhere~\cite{Puebla}
(see also Ref.~\cite{GHS}).
Halzen~\cite{Halzen} has reviewed the
subject of high energy neutrino astronomy at this conference.
Detectors are operating or proposed for deep underwater 
sites and deep in the Antarctic ice.  In addition
to such dedicated neutrino telescopes~\cite{Alvarez}, the
largest air shower detectors
may also have significant aperture 
for neutrinos with $E>10^{19}$~eV~\cite{Cronin},
such as might be radiated from topological defects~\cite{Bhat}.

In the long run, the goal would be to identify specific sources of
cosmic rays, both galactic and extragalactic,
and to see photons, neutrinos and even gravitational
waves that might be associated with them.

\noindent
{ACKNOWLEDGMENTS}.  I am grateful to Ralph Engel, Chuck Smith, Todor
Stanev and Gary Zank for helpful discussions during the preparation
of this talk.


\begin{thebibliography}{99}
\bibitem {Rufolo} David Ruffolo, in {\it Invited, Rapporteur and Highlight
Papers of the 25th International Cosmic Ray Conference} (World Scientific,
ed. M.S. Potgieter, B.C. Raubenheimer, D.J. van der Walt, 1997) p. 109.
\bibitem {JRyan} J.M. Ryan, J.A. Lockwood \& H. Debrunner,
Space Science Reviews 93 (2000) 31.
\bibitem {Ranketal} G. Rank {\it et al.}, Proc. 25th Int.
Cosmic Ray Conf. (Durban, 1997) vol. 1, p. 1.
\bibitem {ACE} R.L. Tokar {\it et al.}, J. Geophys. Res. 105 (2000) 7521.
\bibitem{anomalous} {\it Acceleration and Transport of Energetic
Particles Observed in the Heliosphere} (A.I.P. Conf. Proc. \#528, 2000,
ed. R.A. Mewaldt {\it et al.}) pp. 293-332.
\bibitem{Lozinskaya} {\it Supernovae and Stellar Wind 
in the Interstellar Medium}, Tatiana A. Lozinskaya (American Institute
of Physics, 1992).
\bibitem{Berezhko} E.G. Berezhko \& H.J. V\"{o}lk, 
Astron. \& Astrophys. 357 (2000) 283.
\bibitem {Axford} W.I. Axford, Ap.J. Supplement 90 (1994) 937.
\bibitem {Klepach} E.G. Klepach, V.S. Ptuskin \& V.N. Zirakashvili,
Astroparticle Physics 13 (2000) 161.
\bibitem {Grigorov} N.L. Grigorov {\it et al.} Yad. Fiz. 11 (1970) 1058
and Proc. 12th Int. Cosmic Ray Conf. (Hobart) vol. 2 (1971) 206.
\bibitem {JACEE} K. Asakimori {\it et al.}, Proc. 23rd Int. Cosmic Ray
Conf. (Calgary) vol. 2 (1993) 25.
\bibitem {Akeno} M. Nagano {\it et al.} J. Phys. G10 (1984) 1295.
\bibitem {Dani} T.V. Danilova {\it et al.}, Proc. 15th Int. Cosmic Ray
Conf. (Plovdiv) vol. 8 (1997) 129.
\bibitem {Fomin} Yu. A. Fomin {\it et al.}, Proc. 22nd Int. Cosmic Ray Conf.
(Dublin) vol. 2 (1991) 85.
\bibitem {Tibet} M. Amenomori {\it et al.}, Astrophys. J. 461 (1996) 408.
\bibitem {Glas} M.A.K.  Glasmacher {\it et al.}
Astropart. Phys. 10 (1999) 291.
\bibitem {DICE} S.P. Swordy and D.B. Kieda, Astropart. Phys. 13 (2000) 137.
\bibitem {HEGRA} F. Arqueros {\it et al.}, Astron. Astrophys. 359 (2000) 682.
\bibitem {CasaBlanca} J.W. Fowler {\it et al.}, astro-ph/0003190
(submitted to Astroparticle Physics).
\bibitem {Kascade} R. Glastetter {\it et al.}, Proc. 26th Int. Cosmic
Ray Conf. (Salt Lake City, 1999) vol. 1, p. 222.
\bibitem {LEAP} E.S. Seo {\it et al.}, Ap.J. 378 (1991) 763.
\bibitem {CAPRICE} M. Boezio {\it et al.} Ap.J. 518 (1999) 457.
\bibitem {MASS91} R. Bellotti {\it et al.}, Phys. Rev. D60 (1999) 052002.
\bibitem {IMAX} W. Menn, {\it et al.}, Proc. 25th Int. Cosmic Ray
Conf. (Durban) vol. 3 (1997) 409.
\bibitem {BESS98} T. Sanuki {\it et al.}, astro/ph-0002481
\bibitem {AMSprotons} J. Alcaraz {\it et al.}, Physics Letters B 490 (2000) 27.
\bibitem {Ryan} M.J. Ryan, J.F. Ormes \& V.K. Balasubrahmanyan, Phys. Rev.
Letters 28 (1972) 985 \& E1497.
\bibitem {modulation} L.J. Gleeson \& W.I. Axford, Ap. J. 154 (1968) 1011.
\bibitem {Ginzburg} {\it The Origin of Cosmic Rays},
V.L. Ginzburg \& S.I. Syrovatskii (Pergamon Press, 1964).
\bibitem {Ptuskin} V.S. Ptuskin, A.I.P. Conf. Proc. \#528 (2000) p. 390.
(See Ref.~\cite{anomalous} above.)
\bibitem {Swordy} S. Swordy {\it et al.}, Ap.J. 403 (1993) 658.  See also
S. Swordy, A.I.P. Conf. Proc. \#528 (2000) p. 371.
(See Ref.~\cite{anomalous} above.)
\bibitem {Garcia-Munoz} M. Garcia-Munoz {\it et al.} Ap.J. Supplement
64 (1987) 269.
\bibitem {BESS} S. Orito {\it et al.}, Phys. Rev. Letters 84 (2000) 1078.
\bibitem {Bieberetal} J. Bieber {\it et al.},
Phys. Rev. Letters 83 (1999) 674.
\bibitem {Hunter} S.D. Hunter, Ap.J. 481 (1997) 205.
\bibitem {Cowsik} R. Cowsik in {\it Non-Solar Gamma Rays},
(Pergamon Press, 1979, ed. R. Cowsik \& R.W. Wills) p. 205.
\bibitem {Berezhko2} Berezhko \& V\"{o}lk, Ap.J. 540 (2000) 923.
\bibitem {Drury} L. O'C. Drury, F.A. Aharonian \& H.J. V\"{o}lk,
Astron. Astrophys. 287 (1994) 959.
\bibitem {Esposito} J.A. Esposito, S.D. Hunter, G. Kanbach \& P. Sreekumar,
Ap.J. 461 (1996) 820.
\bibitem {Buckley} J.H. Buckley {\it et al.}, Astron. Astrophys.
329 (1998) 639.
\bibitem {Berezhko3} E.G. Berezhko \& H.J. V\"{o}lk, Astroparticle
Physics 7 (1997) 183 and 14 (2000) 201.
\bibitem {Gaisseretal} T.K. Gaisser, R.J. Protheroe \& Todor Stanev,
Ap.J. 492 (1998) 219.
\bibitem {SN1006} T. Tanimori {\it et al.}, Ap.J. 497 (1998) L25.
\bibitem {CasA} H. V\"{o}lk in {\it GeV-TeV Gamma Ray Astrophysics
Workshop}  (A.I.P. Conf. Proc. 
\#515, 2000, ed. B.L. Dingus, M.H. Salamon \& D.B. Kieda) p. 197.
\bibitem {Pohl}   M. Pohl, Astron. Astrophys. 307 (1996) L57.
\bibitem {Mast}  A. Mastichiadis \& O.C. de Jager, Astron. Astrophys.
311 (1996) L5.
\bibitem {Baring} M.G. Baring in {\it GeV-TeV Gamma Ray Astrophysics
Workshop} (A.I.P. Conf. Proc.
\#515, 2000, ed. B.L. Dingus, M.H. Salamon \& D.B. Kieda) p. 173.
%\bibitem{Peters} knee as more rapid escape
\bibitem {Pt1} V.S. Ptuskin, H.J. V\"{o}lk, V.N. Zirakashvili \& D.
Breitschwerdt, Astron. Astrophys. 321 (1997) 434.
\bibitem {Erlykin} A.D. Erlykin, M. Lipski \& A.W. Wolfendale,
Astroparticle Physics 8 (1998) 283.
\bibitem {Clay} R.W. Clay, M.A. McDonough \& A.G.K. Smith,
Proc. 25th Int. Cosmic Ray Conf. (Durban, 1997) vol. 4, p. 185.
\bibitem {Munakata} K. Munakata {\it et al.}, Phys. Rev. D 56 (1997) 23.
\bibitem {SeoPtuskin} E.S. Seo \& V.S. Ptuskin, Ap.J. 431 (1994) 705.
\bibitem {Simon} M. Simon, W. Heinrich \& K.D. Mathis, Ap.J. 300 (1986) 32
and U. Heinbach \& M. Simon, Ap.J. 441 (1995) 209.
\bibitem {Swordy2} S.P. Swordy {\it et al.}, Ap.J. 349 (1990) 625.
\bibitem {Cacti} S.M. Paling {\it et al.}, Proc. 25th Int. Cosmic Ray
Conf. (Durban, 1997) vol. 5, p. 253.
\bibitem {Vulcan} J.E. Dickinson {\it et al.}, Proc. 26th Int. Cosmic Ray
Conf. (Salt Lake City, 1999) vol. 3. p. 136.
\bibitem {Yakutsk} M.N. Dyakonov {\it et al.}, Proc. 23rd Int. Cosmic Ray
Conf. (Calgary, 1993) vol. 4, p. 303.
\bibitem {Birdetal} D.J. Bird {\it et al.}, Phys. Rev. Letters 71 (1993)
3401.
\bibitem {MIAHiRes} T. Abu--Zayyad {\it et al.}, Phys. Rev. Letters
84 (2000) 4276.
\bibitem {Hinton} J.A. Hinton {\it et al.}, Proc. 26th Int. Cosmic Ray
Conf. (Salt Lake City) vol. 3 (1999) p. 288.
\bibitem {SIBYLL21} R. Engel, T.K. Gaisser, P. Lipari \& T. Stanev,
Proc. 26th Int. Cosmic Ray Conf. (Salt Lake City, 1999) vol. 1, p. 415.
\bibitem {QGSjet} N.N. Kalmykov, S. Ostapchenko \& A.I. Pavlov, Nucl. Phys.
B (Proc. Suppl.) 52B (1997) 17. 
\bibitem {Pryke}  C.L. Pryke, astro-ph/0003442, to be published in
Astroparticle Physics.  The curves for SIBYLL21 and QGSjet protons
have been calculated by R. Engel (private communication).
\bibitem {RJP} R.J. Protheroe \& A.P. Szabo, Phys. Rev. Letters 69
(1992) 2885.
\bibitem {Teshima} N. Hayashida {\it et al.}, Astroparticle Physics
10 (1999) 303.
\bibitem {FEanisotropy} D.J. Bird {\it et al.} astro-ph/9806096.
\bibitem {Clay2} R.W. Clay, B.R. Dawson, J. Bowen \& M. Debes,
 Astroparticle Physics 12 (2000) 249.
\bibitem {AkenoComp} N. Hayashida {\it et al.}, J. Phys. G 21 (1995) 1101.
\bibitem {Dawson} B.R. Dawson, R. Meyhandan \& K.M. Simpson,
Astroparticle Physics 9 (1998) 331.
\bibitem {FE} D.J. Bird {\it et al.}, Astrophys. J. 424 (1994) 491.
\bibitem {HP} M.A. Lawrence, R.J.O. Reid \& A.A. Watson, J. Phys. G17
(1991) 733.
\bibitem {AGASA} M. Takeda {\it et al.}, Phys. Rev. Lett. 81 (1998) 1163.
\bibitem {Yakutsk2} M.I. Pravdin {\it et al.}, Proc. 26th Int. Cosmic
Ray Conf. (Salt Lake City, 1999) vol. 3, p. 292.
\bibitem {Puebla} T.G. Gaisser, Proc. of the International
Workshop on Observing Ultra-High Energy Cosmic Rays From Space and Earth,
(Metepec, Puebla, Mexico, 2000) to be published by A.I.P.
\bibitem {Achterberg} J.G. Kirk, A.W. Guthmann, Y.A. Gallant \& A. Achterberg,
Ap.J. 542 (2000) 235.  See also A. Achterberg, this conference. 
\bibitem {Ostrowski} J. Bednarz \& M. Ostrowski, Phys. Rev. Letters 80
(1998) 3911.
\bibitem {Vietri} M. Vietri, Proc. 9th Marcel Grossmann Conference,
Rome, July 2000.
\bibitem {TSS} T. Stanev {\it et al.}, Phys. Rev. D 62 (2000) 093005.
\bibitem {Grigorieva} V.S. Berezinsky \& S.I. Grigorieva,
Astron. \& Astrophys., 199 (1988) 1.
\bibitem {BW} J.N. Bahcall \& E. Waxman, astro-ph/9912326v2 (2000).
\bibitem {Blasi} M. Blanton, P. Blasi \&  A.V. Olinto,
astro-ph/0009466, submitted to Astroparticle Physics.
\bibitem {dalpino} E.M. de Gouveia Dal Pino |7 A. Lazarian,
astro-ph/0002155.
\bibitem {Olinto} P. Blasi, R.I. Epstein \& A.V. Olinto, astro-ph/9912240.
\bibitem {TD} P. Bhattacharjee, C.T. Hill \& D.N. Schramm, Phys. Rev.
Letters 69 (1992) 567.
\bibitem {Bhat} P. Bhattacharjee \& G. Sigl, Physics Reports 327 (2000) 109.
\bibitem {Berezinsky} V.S. Berezinsky, M Kachelrie\ss \& A. Vilenkin,
Phys. Rev. Letters 79 (1997) 4302.
\bibitem {ProtherStanev} R.J. Protheroe \& T. Stanev, Phys. Rev. Letters
77 (1996) 3708 (and Erratum, 78 (1997) 3420).
\bibitem {Lee} G. Sigl, S. Lee, P. Bhattacharjee \& S. Yoshida, Phys.
Rev. D 59 (1998) 043504.
\bibitem {HSV} F. Halzen, R.A. V'{a}zquez, T. Stanev \& H.P. Vankov,
Astroparticle Physics 3 (1995) 151.
\bibitem {Ave} M. Ave {\it et al.}, Phys. Rev. Letters 85 (2000) 2244.
\bibitem {Ong} R.A. Ong, Physics Reports 305 (1998) 93.
\bibitem {Hoffman} C.M. Hoffman, C. Sinnis, P. Fleury \& M. Punch,
Revs. Mod. Phys. 71 (1999) 897.
\bibitem {AMSproposal} AMS: http://hpl3tri1.cern.ch/AMS/ams\_homepage.html
\bibitem {PAMELA} PAMELA: http://www.particle.kth.se/group\_docs/astro/research/PAMELA.html
\bibitem {Bergstrom} L. Bergstrom, J. Edsj\"{o} \& P. Ullio,
astro-ph/9902012.
\bibitem {ACCESS} ACCESS: http://www701.gsfc.nasa.gov/access/access.htm
\bibitem {HiRes} HiRes Fly's Eye: http://www.cosmic-ray.org/
\bibitem {Auger} Auger: http://www.auger.org/
\bibitem {TA} Telescope Array: http://www-ta.icrr.u-tokyo.ac.jp/
\bibitem {EUSO} EUSO: http://www.ifcai.pa.cnr.it/Ifcai/euso.html
\bibitem {OWL} OWL: http://owl.gsfc.nasa.gov/intro.html
\bibitem {GHS} T.K. Gaisser, Francis Halzen \& Todor Stanev, Physics
Reports 258 (1995) 173.
\bibitem {Halzen} Francis Halzen, this conference.
\bibitem {Alvarez} J. Alvarez-Mu\~{n}iz \& F. Halzen, astro-ph/0007329~v2.
\bibitem {Cronin} K.S. Capelle, J.W. Cronin, G. Parente \& E. Zas,
Astroparticle Physics 8 (1998) 321.
\end{thebibliography}
\end{document}